# High-Speed Multi-Dimensional Optical Field Measurement via MMF-MCF Spatial-Temporal Mapping Architecture

Yuxuan Xiong, Jingze Liu, Junjie Qiu, Zhuyixiao Liu, Zheng Gao, Hao Wu, and Ming Tang, *Member, IEEE*

***Abstract*—Wavelength and state of polarization constitute fundamental dimensions of optical fields. While simultaneous quantification of these parameters is critical, existing methodologies often lack the speed required for real-time analysis. Here, we present a compact high-dimensional optical field analyzer employing a discrete spatiotemporal sampling architecture based on multimode and multicore fibers. An optical delay line array maps spatial speckle patterns into serial pulse sequences and facilitates efficient single-pixel detection. Leveraging a residual multilayer perceptron network, the system attains a wavelength mean absolute error of 0.25 pm and a polarization resolution of 0.2015 (in normalized Stokes space). Analysis of the spatial sampling density reveals that 5-6 sampling points are required to balance measurement rate and accuracy. Notably, the system exhibits isotropic fault tolerance against single-core failures. This confirms that optical field information is redundantly encoded across the entire fiber cross-section rather than localized in specific channels. This framework provides a solution for multiparameter decoupling under severe spatial downsampling and useful insights for the design of next-generation high-speed and robust all-fiber analysis systems.**

## I. INTRODUCTION

SIMULTANEOUS acquisition of high-dimensional optical information, particularly wavelength and state of polarization (SOP), constitutes a critical requirement for applications including optical communication [1-4], material characterization [5,6], and biomedical imaging [7,8]. Conventional approaches typically employ cascaded discrete components such as diffraction gratings and rotating waveplates. However, these systems suffer from substantial physical footprints and demand precise mechanical alignment.

Nanophotonic on-chip spectropolarimeters represent a promising avenue for miniaturization and integration. Recent demonstrations include compact plasmonic rainbow chips [9] and tunable liquid crystal metasurfaces [10] for simultaneous spectral and polarization analysis. Furthermore, operational bandwidths have been extended from the visible to the mid-infrared regime utilizing dispersion-assisted [11] and metasurface-integrated graphene photodetectors [12]. Although these architectures exhibit impressive metrics such as 400 pm spectral sensitivity via disordered-guiding chips [13] and real-time imaging rates exceeding 100 FPS [14], a fundamental limitation persists. High-performance integrated devices often necessitate sophisticated nanofabrication. This imposes an inherent trade-off between ultra-high spectral resolution at the sub-picometer scale and fabrication scalability or cost-effectiveness.

As a cost-effective alternative, multimode fiber (MMF) speckle-based architectures have garnered significant interest for high-resolution metrology [15-18]. By treating the fiber as a complex scattering medium, the MMF encodes high-dimensional optical information into spatial speckle via inter-modal interference [19-24]. Based on this mechanism, we previously demonstrated an MMF-based light analyzer [25], and parallel efforts have yielded all-fiber computational spectropolarimeters exhibiting high polarization accuracy [26]. Nevertheless, these speckle-based methodologies predominantly utilize cameras such as CCD or CMOS devices for image acquisition. Consequently, temporal resolution is fundamentally constrained by the frame rates and data-transfer latencies of conventional image sensors. This limitation precludes the observation of transient optical phenomena.

To overcome the speed limitations inherent in image-based acquisition, recent research has pivoted toward sparse detection schemes utilizing single-pixel or low-pixel-count sensors. Advances have demonstrated that spatial downsampling of the speckle field enables high-speed wavemeters [27] while extreme signal compression via quadrant detectors facilitates rapid fiber Bragg grating wavelength interrogation [28]. However, these sparse sampling strategies remain largely restricted to one-dimensional spectral measurements. Advancing from simple wavelength retrieval to the simultaneous demodulation of wavelength and polarization from sparse signals presents a significant challenge.

In this article, we report a high-speed single-shot measurement

This work was supported by Hubei Optical Fundamental Research Center and the National Natural Science Foundation of China under Grants 62225110, 61931010, Hubei Provincial Natural Science Foundation of China (2025AFB008) and the Major Program (JD) of Hubei Province (2023BAA013). (Corresponding author: Ming Tang and Hao Wu)

Yuxuan Xiong, Jingze Liu, Junjie Qiu, Zhuyixiao Liu, Zheng Gao, Hao Wu, and Ming Tang are with Wuhan National Laboratory for Optoelectronics (WNLO) and National Engineering Laboratory for Next Generation Internet Access System, School of Optical and Electronic Information, Huazhong University of Science and Technology, Wuhan, 430074, Hubei, China. Ming Tang is with Optical Fundamental Research Center, Wuhan, 430074, Hubei, China (Email: tangming@mail.hust.edu.cn, wuhaoboom@hust.edu.cn)

framework capable of simultaneously demodulating wavelength and SOP utilizing a spatiotemporal mapping architecture based on MMF and multicore fiber (MCF) as illustrated in Fig. 1. Through the integration of a fiber delay array, the system transforms two-dimensional spatial speckle patterns into a serial sequence of seven temporal pulses to facilitate data acquisition rates reaching 100 MHz. By leveraging a residual multilayer perceptron (ResMLP) network for signal decoding, the device yields a spectral resolution of 2 pm with a mean absolute error (MAE) of 0.25 pm and a polarization resolution of 0.2015 in normalized Stokes space. Beyond performance metrics, we investigate the relationship between spatial sampling density and measurement speed. It demonstrates that the system requires at least 5-6 sampling points to ensure multi-parameter measurement rate and accuracy. Furthermore, the architecture displays isotropic fault tolerance against single-core failures. The results validate the hypothesis that optical field information is redundantly encoded across the fiber cross-section rather than being confined to specific channels. Moreover, this work offers a strategy for multiparameter decoupling under conditions of severe spatial downsampling.

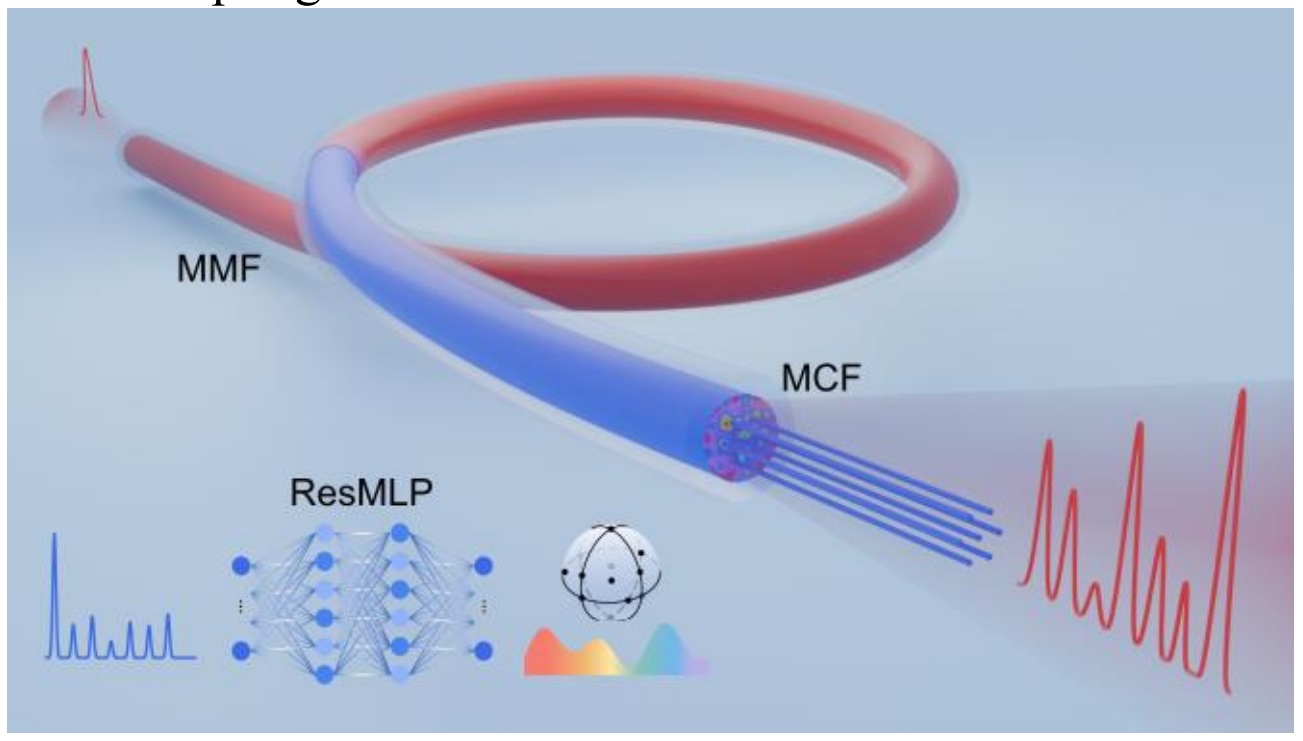

**Fig. 1.** Scheme of the high-dimensional optical field measurement system based on spatial-temporal mapping.
The input optical field excites a speckle pattern in the MMF, which is spatially downsampled by the MCF. A fiber delay line array converts the spatial intensity distribution into a single-shot sparse temporal pulse sequence. The ResMLP is then employed to simultaneously demodulate wavelength and polarization information.

## II. METHOD

### *A. Principle*

MMFs support the propagation of numerous guided modes due to their substantial core diameters. When coherent light is launched into an MMF, inter-modal interference generates a complex speckle pattern at the output facet. The intensity distribution can be described as [29]:

$$I(x,y,L)=\left|\sum_{i=1}^{N}a_i\hat{e}_i(x,y)\exp(j\frac{2\pi}{\lambda}n_{eff,i}L)\right|^2 \quad (1)$$

where $N$ is the number of modes and L is the length of the MMF. In Eq. (1), $a_i$ and $\hat{e}_i$ denote the amplitude and polarization vector of the i-th mode respectively. Consequently, variations in the input wavelength or state of polarization induce a deterministic reconfiguration of the speckle intensity $I(x,y)$ across the end facet [30].

In the proposed architecture, the seven cores of the MCF discretely sample this two-dimensional spatial distribution at specific coordinates $(x_k, y_k)$ where k ranges from 1 to 7. Consequently, the optical power coupled into each core, designated as $P_k$, modulates as a function of the input state. Through parallel-to-serial conversion facilitated by fiber delay lines, these spatial intensity variations transform into distinct peak amplitudes within the temporal domain. As illustrated in Fig. 2, distinct input optical parameters yield unique temporal pulse signatures. This mechanism establishes a deterministic mapping between the high-dimensional input optical field and the one-dimensional output temporal fingerprint. Such a correspondence constitutes the physical foundation for subsequent demodulation via neural networks.

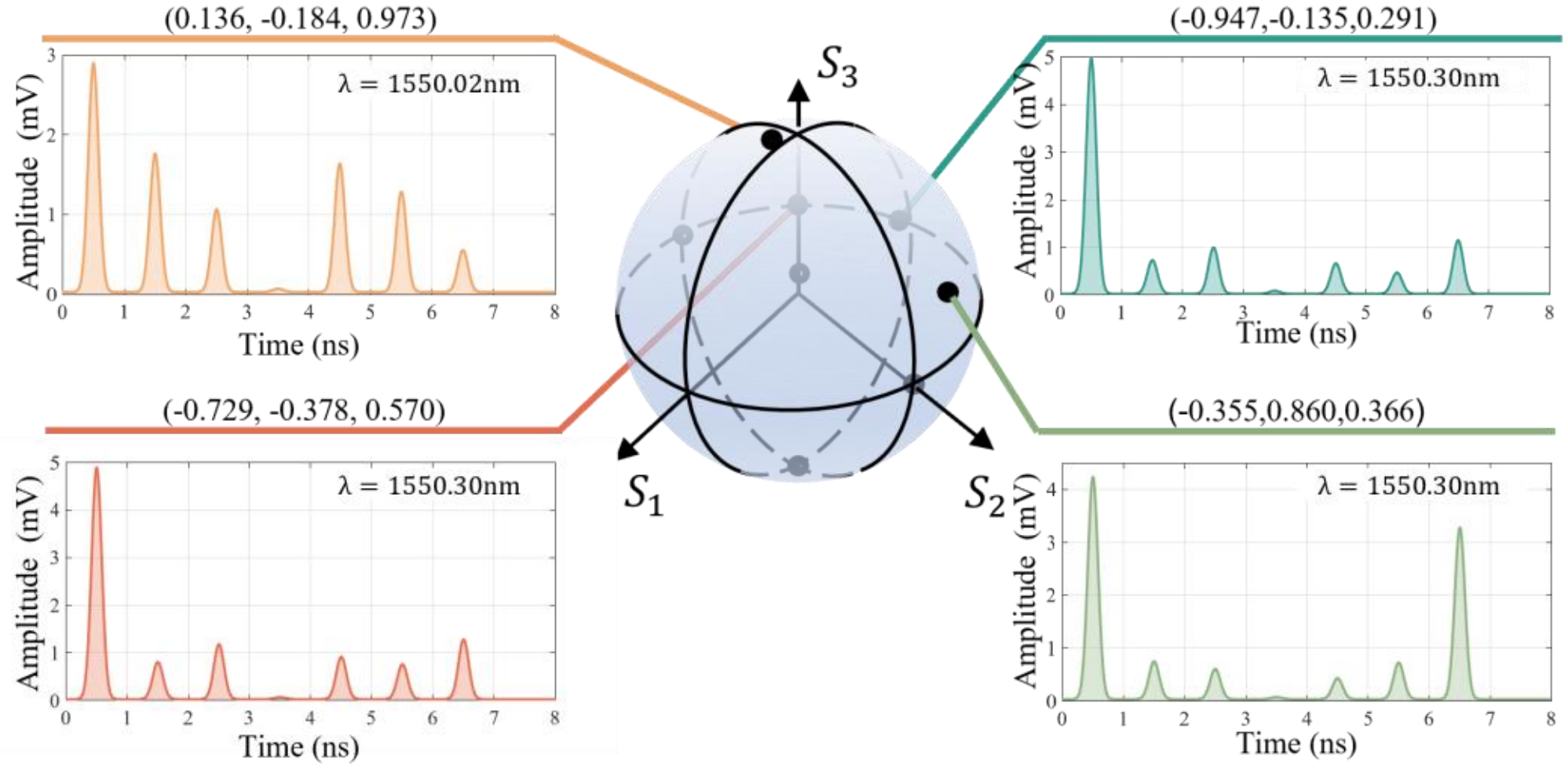

**Fig. 2.** Temporal pulse responses generated by distinct input optical field parameters. The seven-pulse sequences exhibit unique peak amplitude distributions subject to variations in wavelength and SOP. These distinct signatures verify the efficacy of temporal signals for encoding high-dimensional optical information.

*B. Experimental setup*

Fig. 3 illustrates the experimental configuration. A tunable laser initialized at 1549.64 nm functions as the light source. An electro-optic modulator (EOM) driven by an arbitrary waveform generator (AWG) produces optical pulses characterized by a 100 MHz repetition rate and a 0.5 ns pulse width. A motorized polarization controller (MPC) incorporating two quarter-wave plates with a 0.225 degree stepping resolution modulates the input SOP. A 5 m step-index MMF (105/125 μm) is fusion-spliced to a seven-core MCF. A fan-out device spatially segregates the MCF output signals and directs them through a fiber delay line array featuring a 0.2 m length increment to impose a 1 ns temporal separation between adjacent pulses. A 1×8 coupler subsequently combines these channels. Finally, a high-speed photodetector (PD) captures the time-multiplexed seven-pulse sequence for digitization by a digital storage oscilloscope (DSO).

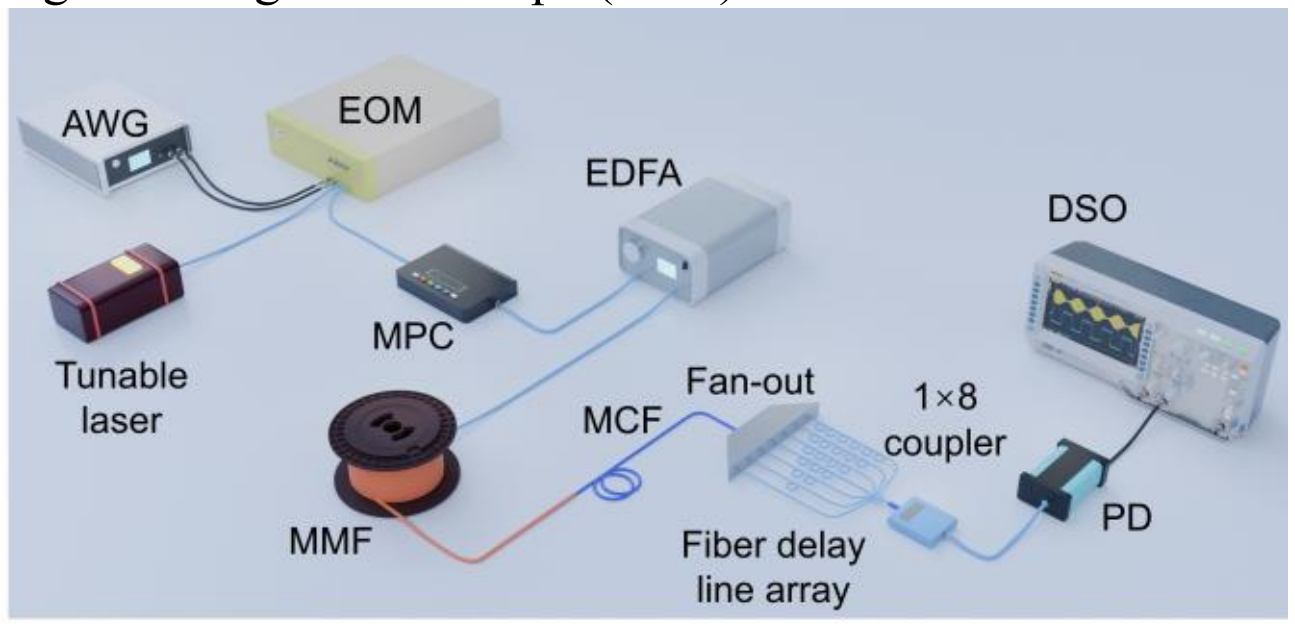

**Fig. 3.** Experimental setup of the MMF-MCF spatial-temporal mapping system.

To assess system performance limits systematically, we acquired comprehensive datasets spanning diverse dynamic ranges. The rotation steps of the waveplates within the controller quantify the polarization dynamic range. Each subset comprises 40 wavelengths × 64 SOPs. The variable $S_{label}$ denotes the corresponding polarization label for the i-th sampled signal. This value is derived via the waveplate transfer function presented in Eq. (2), where $M(\cdot)$ represents the Mueller matrix determined by the rotation angles $\theta_1$ and $\theta_2$ of the respective waveplates, and $S_{in}$ signifies the initial incident SOP emitted by the tunable laser [31,32].

$$S_{label} = M(\theta_1, \theta_2) \cdot S_{in} \quad (2)$$

We acquired datasets characterized by distinct wavelength intervals of 2 pm, 5 pm, 10 pm, and 20 pm. For SOP, the MPC rotation increments are set at 1, 5, 10, 20 and 40 (0.225, 1.125, 2.250, 4.500 and 9.000deg). Two quarter-wave plates rotate from 0 deg to 8 times of interval individually. The different SOP distributions on the Poincare sphere mentioned above are shown in Fig. 4. This sampling strategy facilitates a comprehensive assessment spanning regimes from quasi-static micro-perturbations to large-scale nonlinear dynamics. Calculations determined the standard physical grid steps, defined as the average Euclidean distance between adjacent states, to be 0.0115, 0.0573, 0.1091, 0.2015, and 0.4047 for the 1, 5, 10, 20, and 40-step datasets respectively.

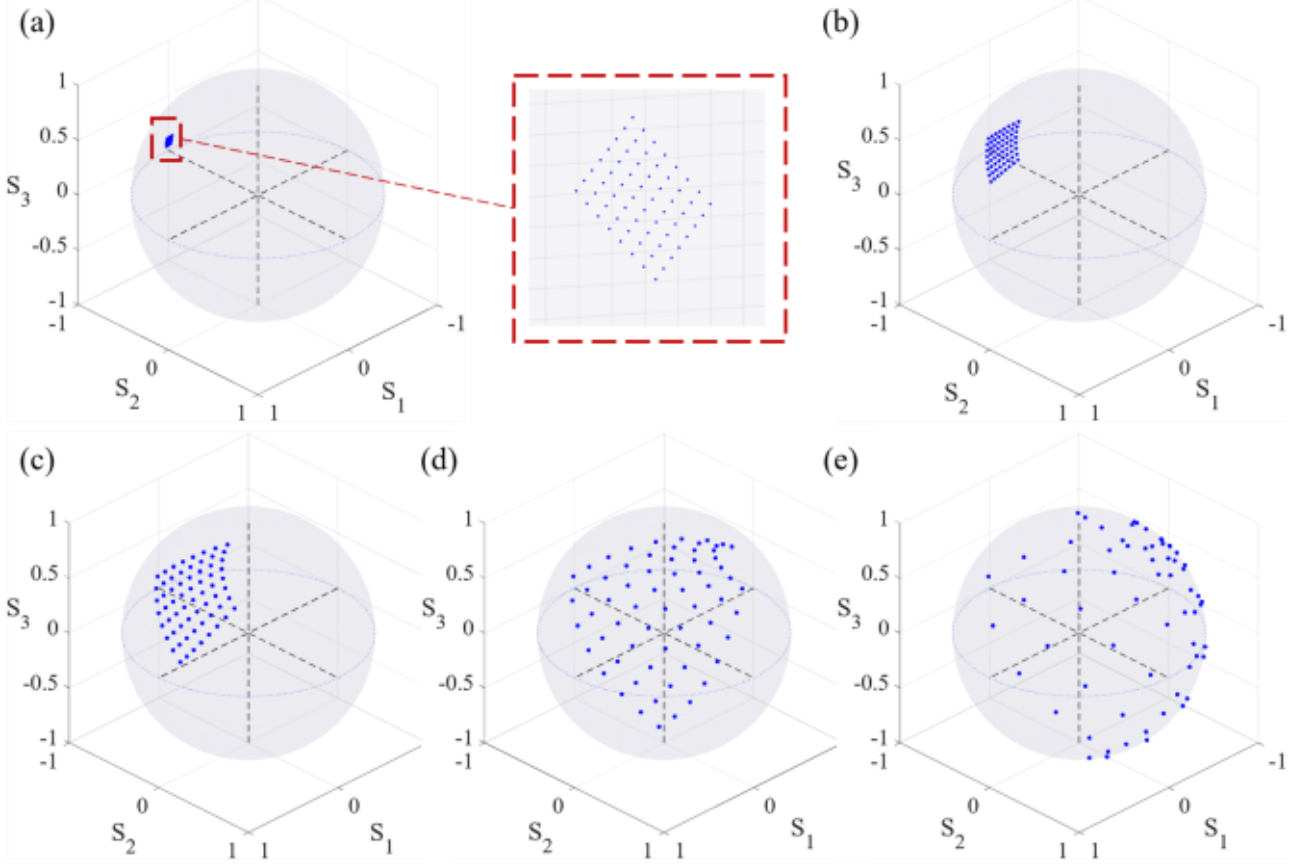

**Fig. 4.** Distribution of SOPs at different densities on the Poincare sphere, covering regions ranging from dense linear regions to sparse nonlinear regions. (a)1 step; (b) 5 steps; (c) 10 steps; (d) 20 steps; (e) 40 steps.

To decode the 1×7 discrete optical intensity pulse signals, we implemented a residual multilayer perceptron (ResMLP) as illustrated in Fig. 5. The architecture initiates with a projection layer that maps the 7-dimensional pulse sequence features onto a 1024-dimensional latent manifold followed by batch normalization (BN) and Gaussian error linear unit (GELU) activation. Deep feature extraction is mediated by a stack of 6 residual blocks that enable the modeling of complex non-linearities while mitigating vanishing gradients. The output stage employs a hierarchical decoding structure that transitions from the 1024-dimensional hidden space to a 256-neuron intermediate layer before yielding the final 4-dimensional regression output comprising the wavelength and three Stokes parameters. We utilized the AdamW optimizer with an initial learning rate of 2×10-4 and a weight decay of 1×10-6 to ensure robust generalization. The training process operated with a batch size of 32 for a maximum of 500 epochs and incorporated a ReduceLROnPlateau scheduler for dynamic learning rate adaptation based on validation loss. To balance spectral and polarization reconstruction, we adopted a custom weighted mean square error (MSE) loss function that assigned a weight ratio of 20 to 1 for the wavelength component relative to the Stokes parameters.

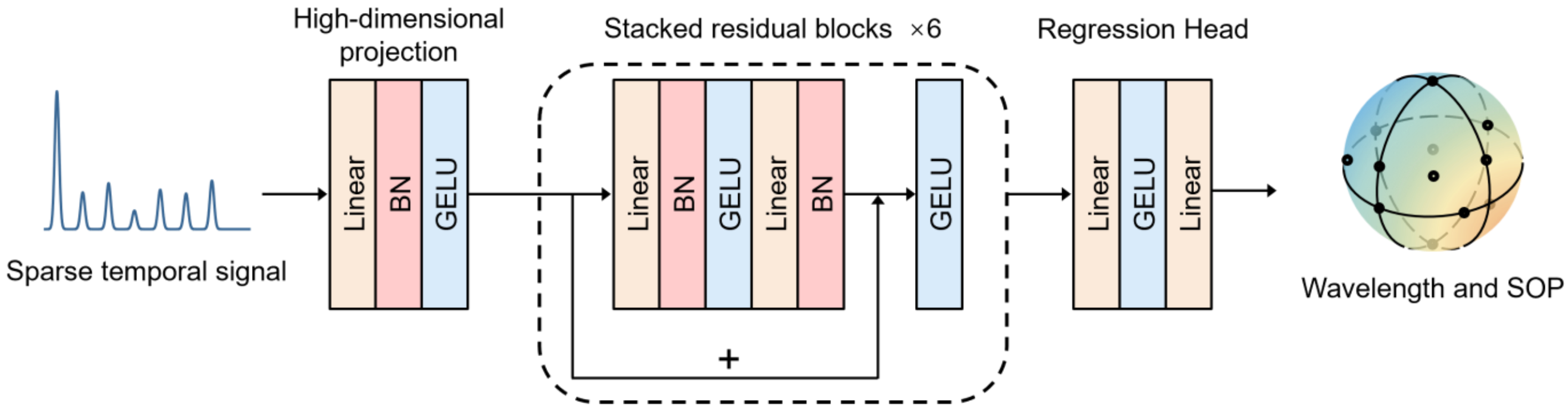

**Fig. 5.** The architecture of the ResMLP. It features high-dimensional projection layers and stacked residual blocks to accurately decouple wavelength and SOP from sparse inputs.

To quantitatively evaluate the demodulation performance, the MAE is selected to characterize wavelength measurement accuracy. Regarding polarization demodulation, the mean Euclidean distance (MED) serves as the metric to quantify physical deviation on the Poincare sphere rigorously as expressed in Eq. 3:

$$\mathrm{MED} = \frac{1}{N_{test}} \sum_{k=1}^{N_{test}} \sqrt{\sum_{j=1}^{3} \left( \hat{S}_{j,act}^{(k)} - S_{j,pre}^{(k)} \right)^2} \tag{3}$$

where $N_{test}$ is the total number of samples in the test set, and $\hat{S}_j^{(k)}$ represents the $j$-th predicted Stokes component for the $k$-th sample. This geometric metric enables a direct comparison between prediction error and the physical grid step of the SOPs and thereby functions as the definitive criterion for determining resolution.

## III. Results and Discussion

### A. System Characterization

To evaluate the precision limits and dynamic range, the ResMLP network was trained on comprehensive datasets ranging from fine perturbation regimes to large non-linear evolution regimes. These datasets included spectral intervals from 2 pm to 20 pm and SOP intervals from 1 step to 40 steps. The data was partitioned into training, validation, and testing sets with an 8:1:1 ratio. Fig. 6 illustrates the decoupling performance under the standard full sampling configuration.

Fig. 6(a) quantifies the spectral reconstruction capabilities. Within the fine-resolution dataset characterized by 2 pm steps, the wavelength MAE attained a minimum of 0.2558 pm. This result exhibits the super-resolution capability of the neural network to interpolate spectral features and resolve minor wavelength shifts effectively. Furthermore, robust tracking accuracy with an error remaining below 2.1 pm persists even for coarser datasets featuring 20 pm intervals and 800 pm bandwidths. This validates system applicability for high dynamic range measurement tasks.

The SOP reconstruction performance is evaluated by the MED on the Poincare sphere in Fig. 6(b). As indicated by the vertical line and gray shaded region, reconstruction uncertainty aligns with physical state evolution for polarization steps below approximately 20 which corresponds to a physical distance of roughly 0.2015. This region delineates the lower bound of detectable polarization changes governed by the system noise floor. Beyond this threshold, the signal dominates the error and facilitates precise analysis.

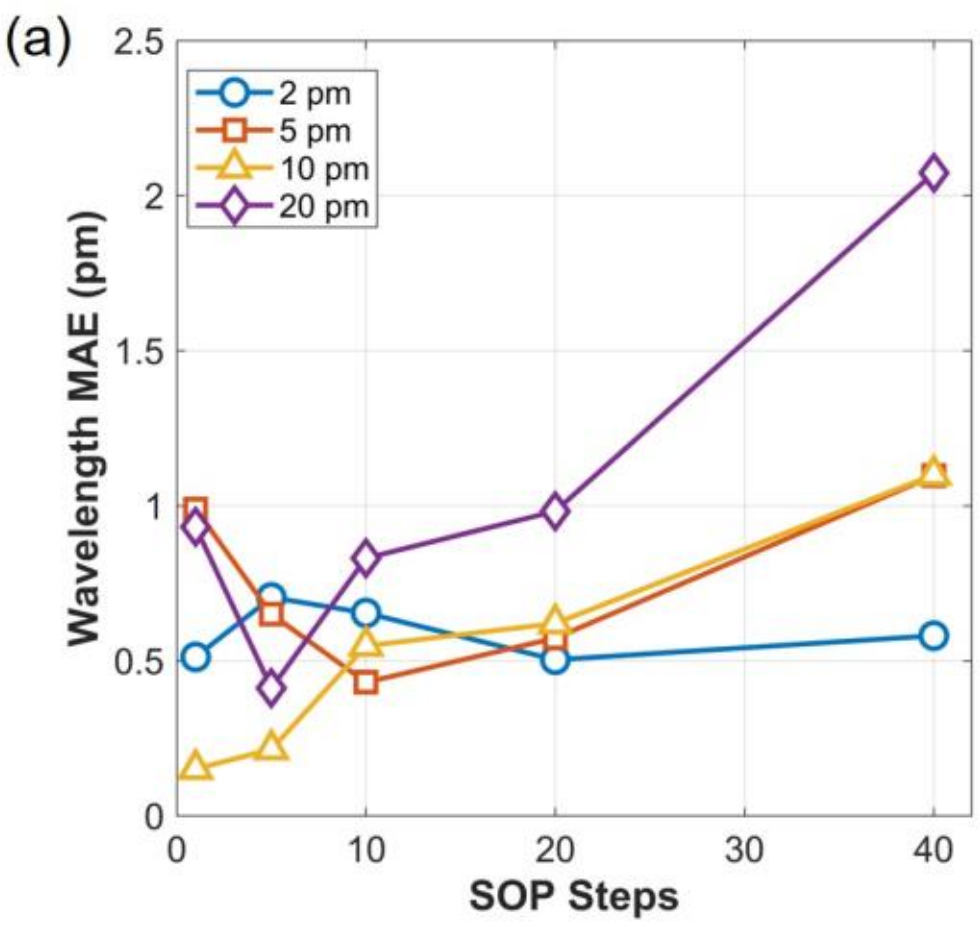

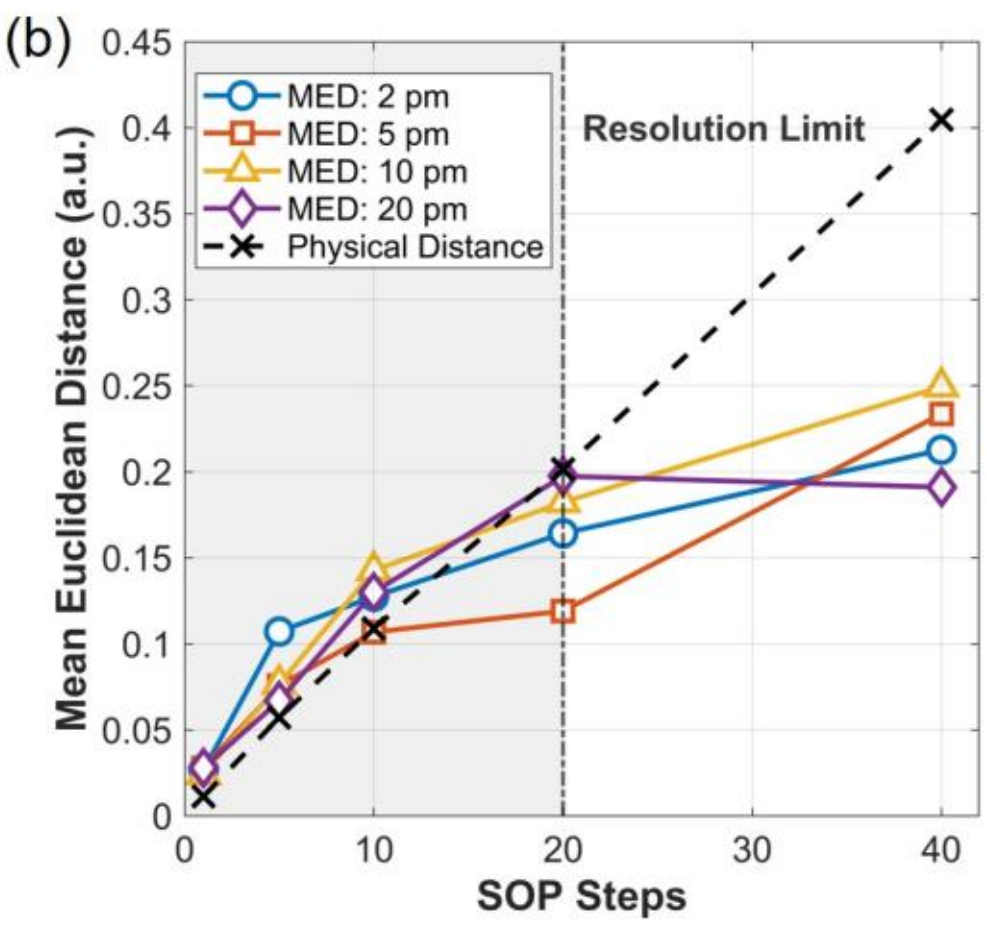

**Fig. 6.** Performance of the ResMLP-based analyzer under the standard full-sampling configuration. (a) Spectral measurement accuracy quantified by the MAE across different SOP steps. (b) Polarization performance evaluated by the MED on the Poincare sphere.

The system maintains high precision despite significant spatial downsampling. Unlike full-field camera-based spectrometers, this architecture inherently performs spatial compression. Given the 105 μm MMF core diameter and seven 10 μm MCF cores, the effective spatial sampling ratio is limited to 6.3%. Consequently, the physical sampling process rejects over 93% of the near-field speckle information. The attainment of picometer-level spectral resolution and robust polarization decoding underscores the superior information efficiency of the multicore fiber-based discrete sampling strategy. Table 1 benchmarks this work against existing representative light field detection schemes. While approaches based on cameras, such as Refs [13, 26], excel in polarization accuracy due to full spatial sampling, they remain constrained by readout and data transmission bottlenecks. Accordingly, their measurement speeds rarely exceed the kHz or Hz regime as evidenced by the 36 Hz rate reported in Ref [14]. Our system distinguishes itself by simultaneously realizing an ultra-high spectral resolution of 2 pm and full-Stokes polarization demodulation while maintaining a 100 MHz repetition rate. This performance effectively bridges a critical gap in the field of high-speed and high-precision optical field analysis.

TABLE I
COMPARISON OF PERFORMANCE WITH DIFFERENT LIGHT FIELD DETECTION SCHEMES

| Refs | Bandwidth & Resolution | Polarization ability | SOP MED | Measurement Speed |
|---|---|---|---|---|
| 9 | 270 nm<br>2 nm | Linear | 0.0079* | NA |
| 11 | 500 nm<br>2.7nm | Full-Stokes | 0.12* | NA |
| 13 | 20 nm<br>400 pm | Full-Stokes | 0.0209* | NA |
| 14 | 450 nm<br>0.23nm | Full-Stokes | NA | 36 Hz* |
| 25 | 100 pm<br>2 pm | Full-Stokes | 0.0152* | NA |
| 26 | 30 nm<br>100 pm | Full-Stokes | 0.0126* | NA |
| This work | 800 pm<br>2 pm | Full-Stokes | 0.1191 | 100 MHz |

(NA, indicates that no specific metrics are provided in the reference, while * denotes values derived from the mentioned parameters in the reference.)

### *B. Trade-off between Spatial Sampling Density and Reconstruction Accuracy*

Optical delay lines represent a necessary component for serializing parallel spatial signals into a single pixel detector. However, the expanded temporal window fundamentally imposes an upper limit on the measurement rate. The feasibility of hardware simplification was investigated by progressively reducing the number of active cores N from the standard seven core configuration down to a single core. Fig. 7 and Fig. 8 illustrate the evolution of spectral MAE and polarization MED to visualize the impact of information loss on reconstruction fidelity.

As the number of sampling fiber cores scales from 1 to 5, the MAE of the wavelength estimation decreases rapidly, since each additional channel introduces independent spectral information. Beyond five cores, the improvement in wavelength accuracy diminishes and plateaus, indicating asymptotic convergence. Moreover, for SOPs, the number of sampling fiber cores exerts a significant influence on measurement accuracy, exhibiting an approximately linear correlation as the range of the states to be measured increases.

Given the requirement for simultaneous multi-parameter analysis, a minimum of five to six fiber cores is essential to sustain the current level of precision. While further increasing the core count, for instance by replacing a seven-core fiber with a nineteen-core fiber or other space-division multiplexing configurations, can enhance the accuracy of wavelength and polarization state reconstruction, it simultaneously imposes a penalty on the detection speed of the system. This inverse relationship between spatial sampling density and temporal throughput necessitates a fundamental trade-off during the system design process.

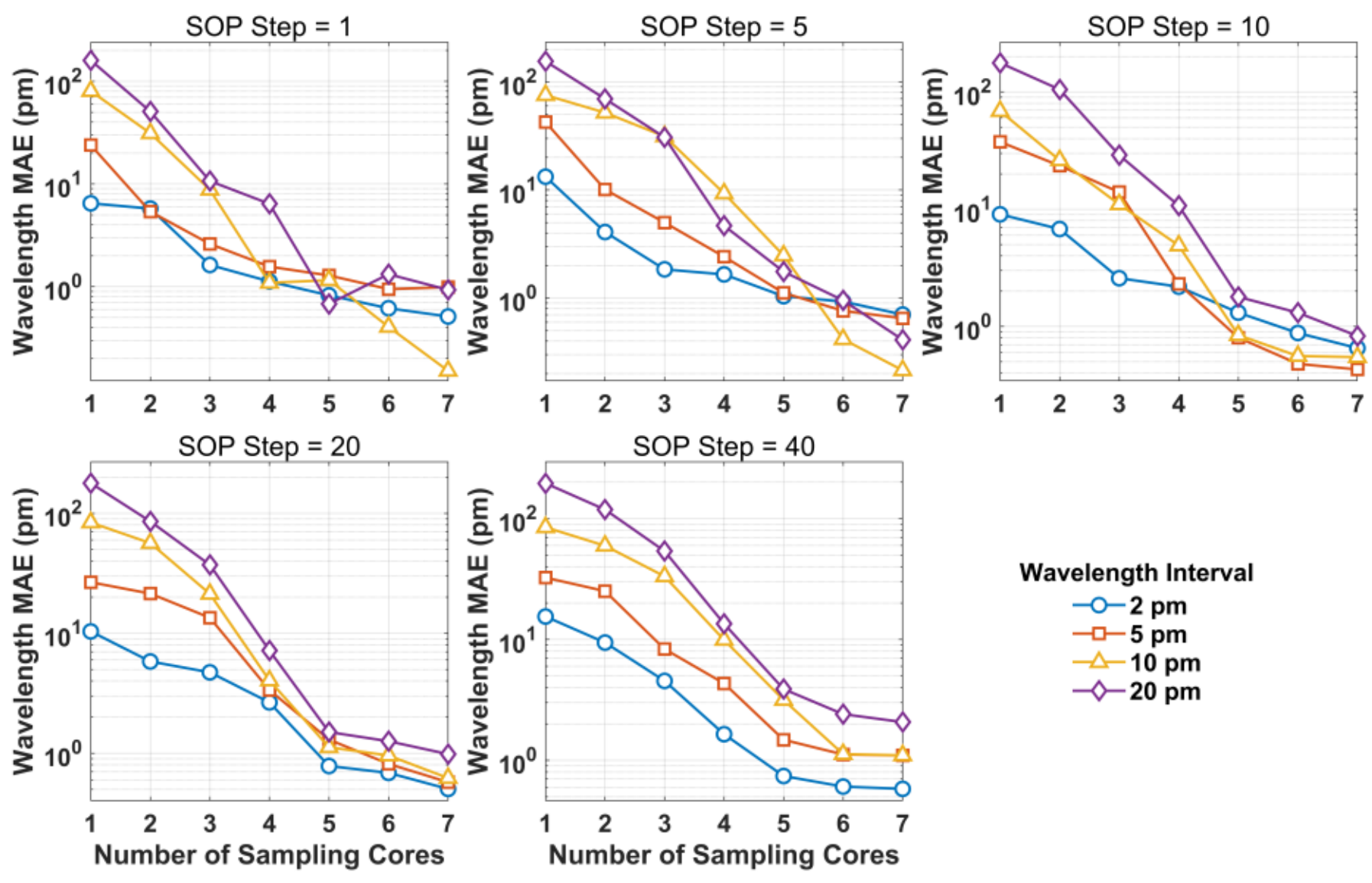

**Fig. 7.** Dependence of spectral reconstruction accuracy on spatial sampling.

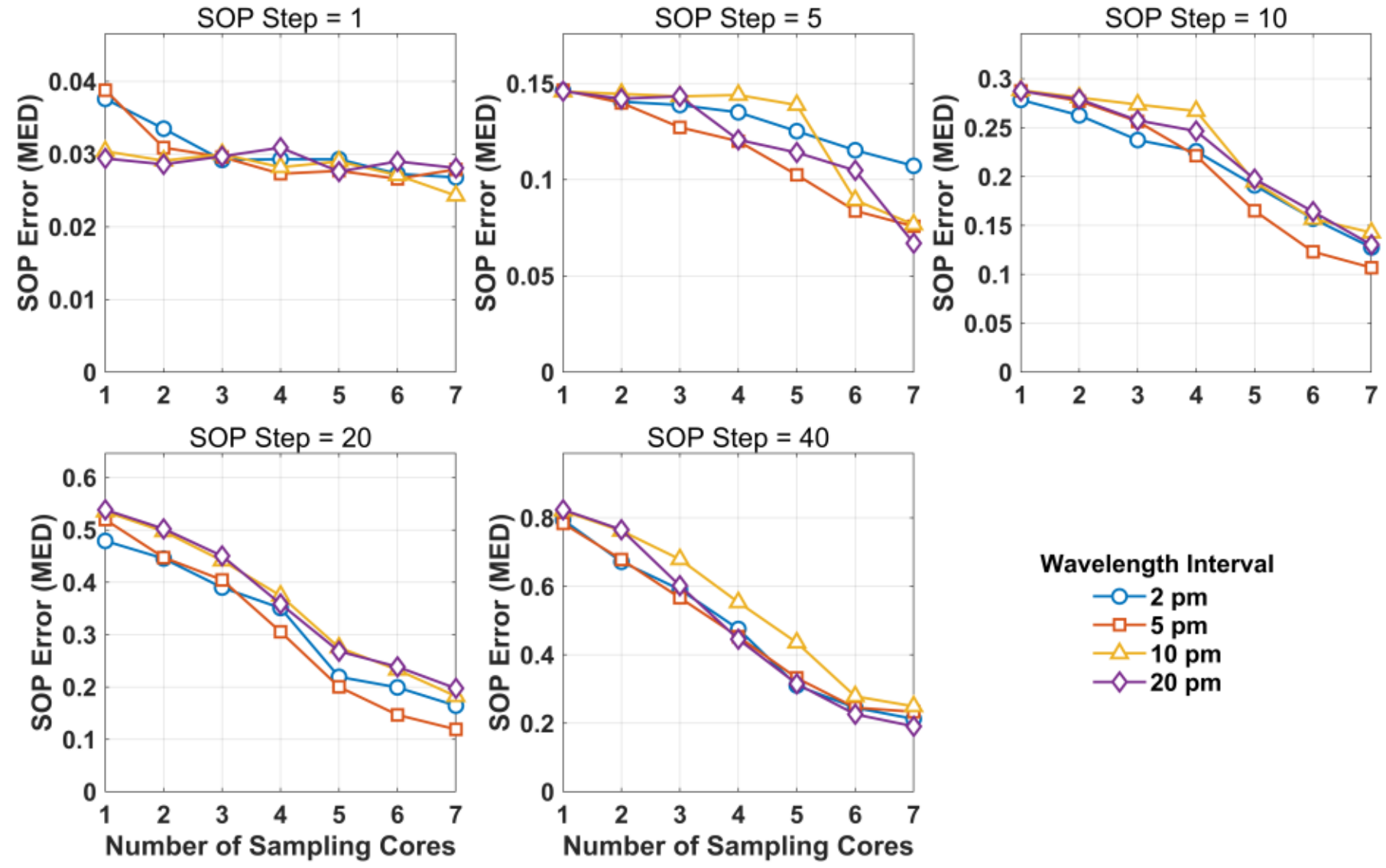

**Fig. 8.** Dependence of SOP analysis accuracy on spatial sampling.

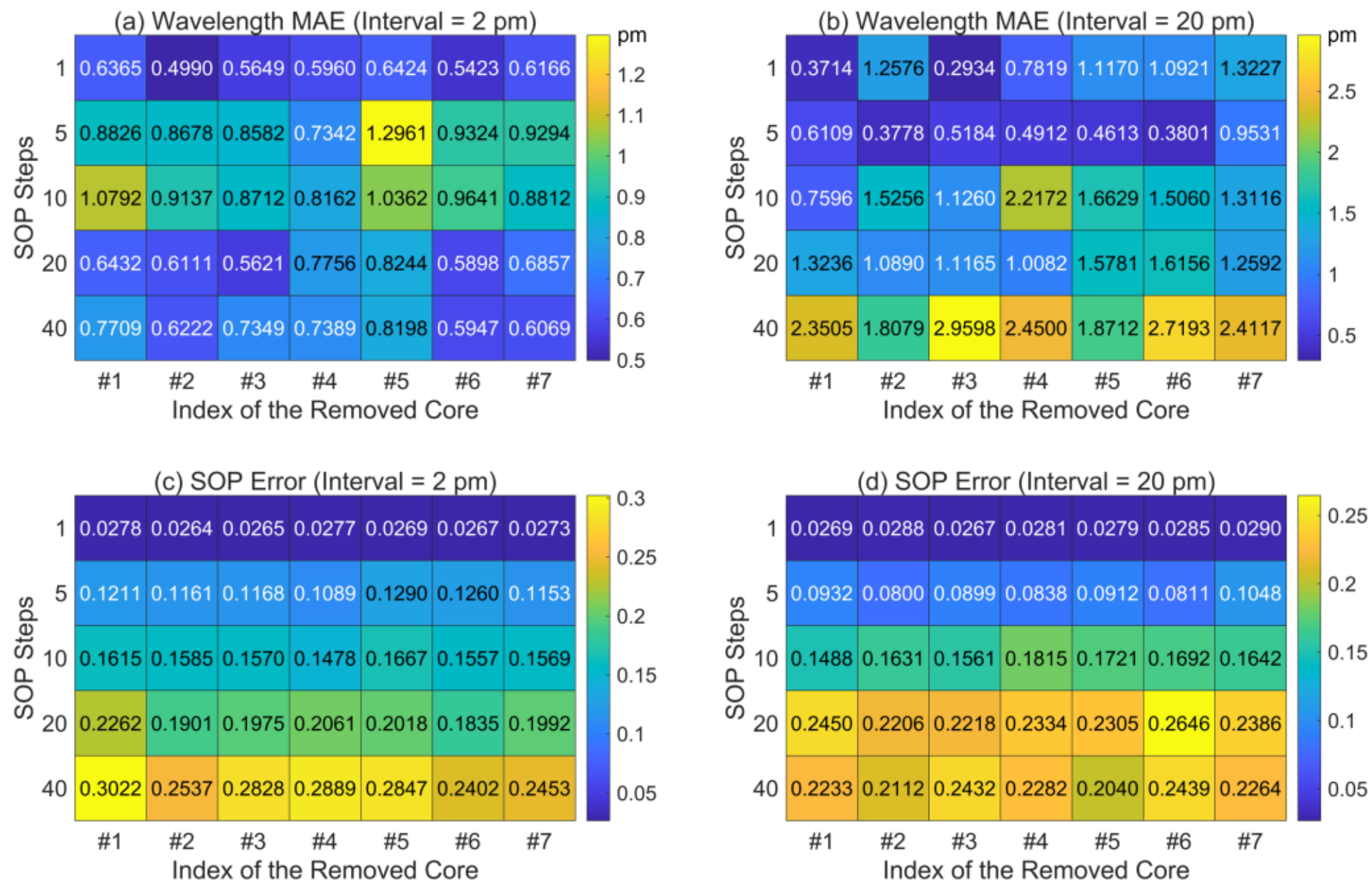

**Fig. 9.** Spatial position dependence and robustness analysis under single-core loss scenarios.
(a-b) Heatmaps of Wavelength MAE for (a) fine-scale (2 pm interval) and (b) large-dynamic-range (20 pm interval) datasets.
(c-d) Heatmaps of SOP MED for (c) 2 pm and (d) 20 pm intervals.

### *C. System Robustness under Single-Core Failure*

The stochastic nature of speckle patterns in MMFs inherently induces variations in spatial intensity and information density. To eliminate potential bias and verify system reliability under partial hardware failure conditions, a systematic single-core masking experiment was conducted. This procedure ensures uniformity in the information retrieval capability. Unlike systems constrained by a dominant central mode, the results indicate that removing any single core results in comparable error levels without significant performance degradation. This that high measurement precision does not originate from a specific core fortuitously capturing a region of high information entropy. The model was trained and tested by iteratively masking the signal from one of the seven cores and relying solely on the remaining six.

The heatmaps in Fig. 9 exhibit a high degree of spatial observation persists even for the central Core 1. Under the complex condition characterized by a 20 pm interval and a 40-step prediction horizon as depicted in Fig. 9(d), the polarization reconstruction error remains stable within a narrow range of 0.20 to 0.24 regardless of which core undergoes masking.

This isotropic fault tolerance confirms that no single core is critical for overall accuracy. Instead, spectral and polarization information is redundantly encoded across the entire speckle pattern. The neural network effectively extracts these distributed features to ensure robust operation. This implies a high tolerance for splicing angles and alignment precision between MMF and MCF, relaxing the requirements for practical implementation and packaging.

## IV. Conclusion

We have proposed and demonstrated a high-dimensional optical field analyzer utilizing a discrete spatiotemporal sampling architecture based on MMF and MCF. The incorporation of optical delay lines translates parallel spatial speckle information into the time domain to facilitate efficient single-pixel detection. Powered by ResMLP, the system delivers a spectral accuracy of 0.25 pm and a polarization resolution of 0.2015. Our analysis of sampling density indicates that at least 5-6 sampling points are required to balance sampling accuracy and measurement rate in multi-parameter analysis. Moreover, the verification of isotropic fault tolerance corroborates that optical field information is redundantly encoded across the fiber cross-section. This feature negates reliance on any specific spatial channel. The proposed methodology realizes high-speed multiparameter decoupling under regimes of severe spatial downsampling and provides valuable insights for the design of next-generation high-speed, robust all-fiber analysis systems.

## Acknowledgment

This work was supported by National Natural Science Foundation of China under Grants 62225110, 61931010; Hubei Provincial Natural Science Foundation of China (2025AFB008) and Major Program (JD) of Hubei Province (2023BAA013).